\documentstyle[floats,aps,prl]{revtex}
\begin{document}
\input epsf
\draft
\twocolumn[\hsize\textwidth\columnwidth\hsize\csname  
@twocolumnfalse\endcsname
\bibliographystyle{}

\title{Cauchy magnetic field component and magnitude distribution studied 
by the zero-field muon spin relaxation technique}
\author{ X. Wan \and W. J. Kossler }
\address{Physics Department, College of William and Mary, Williamsburg, 
VA 23187}
\author{ C. E. Stronach \and D. R. Noakes}
\address{Physics Department, Virginia State University, Petersburg, VA 23806}
\date{\today}
\maketitle

\begin{abstract}

Zero-field muon spin relaxation (ZF-$\mu $SR) data for dilute spin
magnetic
systems have been widely interpreted with what is called a Kubo-Toyabe
form
based on a Lorentzian distribution of local field components. We derive
here
the proper magnetic field \textit{magnitude} distribution using
independent
and uncorrelated \textit{component} distributions. Our result is then
compared to the previously accepted formula for ZF-$\mu $SR. We discuss
the
origins of the magnetic field component and magnitude distributions.
Further
we found that after rescaling the magnetic field, the differences that are
amenable to experimental examination are quite small, although the
interpretations behind them are quite different.

\end{abstract}

\vskip2pc]  
 
\narrowtext 

Zero-field Muon Spin Relaxation (ZF-$\mu $SR)[1] has been used long and
widely to study the local magnetic field distribution in the rare-earth
metallic alloys, spin glasses, heavy fermion systems, superconducting
compounds and other magnetic systems.[2] It has been believed[3] that in the
case of the dilute limit of sparse magnetic moments, the local field
component distribution is Cauchy or Lorentzian-like:

\begin{equation}
g(B_i)=\frac 1\pi \frac \alpha {\alpha ^2+B_i^2}   \label{e1}
\end{equation}
where $B_i=B_x$, $B_y$ or $B_z$, and $\alpha $ is the half-width at
half-maximum (HWHM).

On the other hand, from studies of systems with randomly distributed dilute
magnetic impurities it was concluded that the internal field \textit{%
magnitude} distribution is expected to be:[4--7]

\begin{equation}
P(\left| \mathbf{B}\right| )=\frac 4\pi \frac{\Gamma \left| \mathbf{B}%
\right| ^2}{(\Gamma ^2+\left| \mathbf{B}\right| ^2)^2}   \label{e2}
\end{equation}
\\Consequently the ZF-$\mu $SR relaxation function takes the Lorentzian
Kubo-Toyabe form

\begin{equation}
G_z(t)=\frac 13+\frac 23(1-\lambda t)e^{-\lambda t}    \label{e3}
\end{equation}  
\\where $\lambda =\gamma _\mu \Gamma $, and $\gamma _\mu $ is the
gyromagnetic ratio of the muon 2$\pi $(13.554 kHz/G).\\

In the following calculation, we will prove that the magnetic field
magnitude distribution corresponding to the component distribution (1) is:

\begin{equation}
P_c(\left| \mathbf{B}\right| )=\frac{12\left| \mathbf{B}\right| \arctan
\sqrt{2(\frac{\left| \mathbf{B}\right| }\alpha )^2+(\frac{\left| \mathbf{B}%
\right| }\alpha )^4}}{\alpha ^2\pi ^2[(\frac{\left| \mathbf{B}\right| }%
\alpha )^2+3]\sqrt{2+(\frac{\left| \mathbf{B}\right| }\alpha )^2}} 
\label{e4}
\end{equation}
\\

If the components of the magnetic field are \textit{independent} and \textit{%
uncorrelated} \textit{with each other} then the three dimensional
probability distribution p(\textbf{B}) $=g(B_x)g(B_y)g(B_z)$. Note that this
probability distribution is \textbf{not} isotropic. In fact the three
dimensional distribution as a product of individual component distributions
is only isotropic for the case of gaussian individual component
distributions.\\

The magnitude distribution $P_c(\left| \mathbf{B}\right| )$ may be written
as an integral over the p(\textbf{B}):

\begin{eqnarray}
P_c(\left| \mathbf{B}\right| ) & = & \int_0^\pi \int_0^{2\pi }g(\left|
\mathbf{B}%
\right| \sin \theta \cos \varphi )g(\left| \mathbf{B}\right| \sin \theta
\sin \varphi )  \nonumber \\
& & g(\left| \mathbf{B}\right| \cos \theta )\left|
\mathbf{B}%
\right| ^2\sin \theta d\varphi d\theta   \label{e5}
\end{eqnarray}
  
Inserting Eq. (1) into the expression of $P_c(\left| \mathbf{B}\right| )$
above, assuming $u=\frac{\left| \mathbf{B}\right| }\alpha $

\begin{equation}
P_c(\left| \mathbf{B}\right| )d\left| \mathbf{B}\right| =f(u)du 
\label{e6}
\end{equation}
where

\begin{eqnarray}
f(u) & = & \frac 1{\pi ^3}\int_0^\pi \int_0^{2\pi } \nonumber \\ 
& & \frac{u^2\sin \theta d\varphi
 d\theta }{(1+u^2\cos ^2\theta )(1+u^2\sin ^2\theta \cos ^2\varphi
)(1+u^2\sin ^2\theta \sin ^2\varphi )} \nonumber \\
& &   \label{e7}
\end{eqnarray}

In the equation above, we integrate $\varphi $ first. By using the \textit{%
residue theorem}

\begin{eqnarray}
&&\int_0^{2\pi }\frac{d\varphi }{(1+u^2\sin ^2\theta \cos ^2\varphi
)(1+u^2\sin ^2\theta \sin ^2\varphi )}  \nonumber \\
&=&\frac{2\pi }{(1+\frac 12u^2\sin ^2\theta )\sqrt{1+u^2\sin ^2\theta }}
  \label{e8}
\end{eqnarray}

Putting this result back into Eq. (7), we get

\begin{eqnarray}
f(u) &=&\frac{2u^2}{\pi ^2}\int_0^\pi \frac{\sin \theta d\theta }{(1+u^2\cos
^2\theta )(1+\frac 12u^2\sin ^2\theta )\sqrt{1+u^2\sin ^2\theta }}  \nonumber
\\
&=&\frac{4u^2}{\pi ^2}\int_0^1\frac{dt}{(1+u^2t^2)(1+\frac 12u^2-\frac
12u^2t^2)\sqrt{1+u^2-u^2t^2}}\text{ }(\text{where }t=\cos \theta )  \nonumber
\\
&=&\frac{12u\arctan \sqrt{2u^2+u^4}}{\pi ^2(u^2+3)\sqrt{2+u^2}} 
\label{e9}
\end{eqnarray}

By using the identity Eq. (6), we finally get the magnetic field magnitude
distribution Eq. (4). A study of this field distribution discloses two
important features:

a) At low field, the field magnitude distribution, Eq. (9), is asymptotic to
$\frac{4\left| \mathbf{B}\right| ^2}{\alpha ^3\pi ^2}$, i.e. proportional to
$\left| \mathbf{B}\right| ^2$. This is the same asymptotic behavior shown by
Eq. (2), a result expected for a variety of models.[4] However, it should be
noted that it is difficult to obtain a specific heat which is linear in $T$
at low temperatures as observed experimentally[8] with a field magnitude
distribution which is proportional to $\left| \mathbf{B}\right| ^2$ while
using the molecular-field model with the spins of the impurities quantized
along the local field and no overall preferred spin direction.[4] On the
other hand, there exist other physical models[9,10] which would give at low  
field a constant field-magnitude asymptote. These models would yield a
linear low temperature $T$ dependence for the specific heat.

b) It can be determined that the \textit{Wronskian} of Eq. (4) and Eq. (2)
(even the rescaled Eq. (2)) is nonzero which implies that they are indeed
independent i.e. $P_c(\left| \mathbf{B}\right| )\neq C\cdot P(D\cdot \left|
\mathbf{B}\right| )$ where $C$ and $D$ are arbitrary constants.
Nevertheless, it is interesting to compare these two distributions
numerically. If one starts from Eq. (2) and uses $v=\frac{\left| \mathbf{B}%
\right| }\Gamma $, one obtains
\begin{equation}
P(\left| \mathbf{B}\right| )d\left| \mathbf{B}\right| =h(v)dv 
\label{e10}
\end{equation}
where

\begin{equation}
h(v)=\frac 4\pi \frac{v^2}{(1+v^2)^2}   \label{e11}
\end{equation}

$h(v)$ and $f(u)$ have maxima at $v_0=\frac{\left| \mathbf{B}\right| }\Gamma
=1$ and $u_0=\frac{\left| \mathbf{B}\right| }\alpha \simeq 1.476$
respectively. In order to compare these two distributions we let them reach
maximum at the same magnetic field $\left| \mathbf{B}\right| $. Then we  
obtain the following relations:

\begin{equation}
\Gamma =1.476\alpha   \label{e12}
\end{equation}

\begin{equation}
j(u)=\frac 4\pi \frac{(\frac u{1.476})^2}{[1+(\frac u{1.476})^2]^2}\frac
1{1.476}   \label{e13}
\end{equation}
where $j(u)$ is just a rescaled distribution of $h(v)$ for $u=1.476v$.

A plot showing the similarity and difference between  $f(u)$ and $j(u)$ is
shown in Fig. 1.

\begin{figure}[h]
\vskip 1.5in
\centerline{\epsfxsize=2.375in \epsfbox{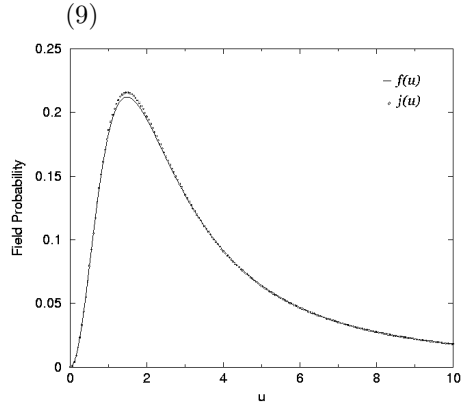}}
\vskip .25in  
\caption{Comparison of the probability distributions of the magnitude
of the magnetic field between $f(u)$ and $j(u)$.}
\end{figure}

It is interesting to study Eq. (2) further. Suppose we divide it by $4\pi   
\left| \mathbf{B}\right| ^2$ to change it to a probability \textit{density}
distribution (we assume here the field is isotropic) and do a double
integral in the Cartesian coordinate system as follows:

\begin{eqnarray}
P_x(B_x) &=&\int_{-\infty }^{+\infty }\int_{-\infty }^{+\infty }\frac 1{\pi
^2}\frac \Gamma {(\Gamma ^2+B_x^2+B_y^2+B_z^2)^2}dB_ydB_z  \nonumber \\
&=&\frac 1\pi \frac \Gamma {\Gamma ^2+B_x^2}   \label{e14}
\end{eqnarray}

If we take $\Gamma =\alpha $, the above equation is exactly Eq. (1). At 
first glance, this is quite puzzling since we started from the same
distribution and obtained a totally different magnitude distribution: Eq.
(4). A careful study of all the derivation procedures above, leads us to
conclude that: there is not a one to one mapping or correspondence between
the component distribution and magnitude distribution; to some extent, the
two seemingly paradoxical results are due to different causalities; the
three component distributions behind the magnitude distribution Eq. (4) are
assumed to be independent and uncorrelated, on the other hand, the component
distribution Eq. (14) arises from the magnitude distribution assuming
isotropy.

It can be seen from figure 1 that although the form of our new result is
analytically quite different from the previous one which is widely used in
ZF-$\mu $SR spectroscopy, numerically they are amazingly close to each
other. As far as the relaxation function is concerned, an experimental 
determination of which field magnitude distribution is correct should be
essentially impossible.

A commonly-used procedure[4] which yields the Lorentzian field magnitude
distribution of Eq. (2) is to assume a RKKY(Ruderman-Kittel-Kasuya-Yosida)
interaction among the dilutely distributed spins and to use the MRF(Mean 
Random Field) approximation. Further this produces the distribution of the
vector field $\mathbf{B}$ (not just the individual component distributions).
Even for an isotropic vector field distribution, it rarely occurs that this
distribution factors into 3 uncorrelated field component distributions. Our
approach to this problem is logically straightforward and independent of any
physical models, the only assumption here is the use of an appropriate field
component distribution.

\bibliography{paper2}
\end{document}